# Electronic transport and specific heat of 1T- VSe$_2$

C S Yadav[1,2] and A K Rastogi[1]

[1]School of Physical Science, Jawaharlal Nehru University, New Delhi-110067, India

[2] Department of Condensed Matter Physics & Material Science, Tata Institute of Fundamental Research, Mumbai- 400005, India

E-mail: csyadav@tifr.res.in

*Abstract*

The results of low temperature thermoelectric power and the specific heat of 1T-VSe$_2$ (Vanadium diselenide) have been reported along with the electrical resistivity, and Hall coefficient of the compound. The Charge Density Wave (CDW) transition is observed near 110K temperature in all these properties. The Thermoelectric power has been measured from 15K to 300K spanning the incommensurate and commensurate CDW regions. We observed a weak anomaly at the CDW transition for the first time in the specific heat of VSe$_2$. The linear temperature dependence of resistivity and thermoelectric power at higher temperatures suggests a normal metallic behavior and electron-phonon scattering above the CDW transition. The positive thermoelectric power and negative Hall coefficient along with strongly temperature dependent behavior in the CDW phase suggest a mixed conduction related to the strongly hybridized s-p-d bands in this compound.

## 1. Introduction

Among the layered compounds of transition metal dichalcogenides, 1T-VSe$_2$ possesses some peculiar properties. The magnetic and thermal properties at low temperatures suggest a narrow band with exchange enhanced susceptibility [1]. But the conductivity behavior and a transition at 110K in 1T-VSe$_2$ have similarity to the CDW transition in more covalent polytypes of 2H-variety of NbSe$_2$, TaS$_2$ and TaSe$_2$. [1,2,3,4,5] Hence it suggests that in this layered compound 110K - transition is relatively insensitive to the Fermi surface topology and nesting vectors. Moreover, the detailed characterization of CDW-related lattice distortion, its phasing in different lattice directions, the location of the onset and commensurate-distortions temperatures $T_0$ and $T_d$ etc., have been controversial in VSe$_2$. The successions of incommensurate superlattice were observed much above 110K in an electron diffraction study [6], and a drop in the magnetic susceptibility with hysteresis much above 110K was also subsequently reported [7]. The Nuclear magnetic resonance (NMR) study of CDW state showed a clear pre-transitional broadening of the $^{51}$V quadrupole satellites and $^{77}$Se resonance line [8, 9]. A detailed account of the lattice distortions observed in different studies has been given by K. Tsutsumi et al and they reported for the first time a new unusual superstructure at the onset temperature 110K that is commensurate (4a×4a) in the plane of layers, but incommensurate perpendicular to it [10,11].

Up to 1.5% excess of V in the van der Waals sites, does not affect the CDW transition. Similar results have also been reported by us where up to 5% intercalation of Fe does not affect the transition [12]. Although the anomalies in resistance and Hall coefficient are quite clear, these anomalies become rather weak in the magnetic susceptibility, and are controversial in the structural properties [1, 2]. On the other hand, before this report, no appreciable anomaly was reported in the specific heat. [13] The application of external pressure enhances the CDW anomalies and transition temperature shifts to higher temperature. This has been explained by Friend and Jerome, as due to the broadening of d-band of V, and the reduction in the coulomb repulsion effects on CDW [14]. The substitutional doping of Nb also induces the same effects, and the magnetic susceptibility results show an increase in transition temperature to 220K for 1.5% of Nb [15].

In the present study, we have investigated the thermoelectric power (TEP), and the Specific heat, of the $VSe_2$. The Thermopower results were found to be very appealing and contrary to earlier report [12], show linear temperature dependence above the CDW transition. A weak anomaly in the specific heat at CDW transition has been observed and the electronic and phonon contribution to specific heat have been reported for the first time. We have further explored the role of the strains/impurities on the electrical resistivity of the compound. The low temperature Hall coefficient behavior is also presented to show the mixed conduction behavior in the compound.

## 2. Experimental details

The compound was prepared by reacting V chips (99.7%) and 5% excess Se granules (99.999) over required composition in a sealed quartz ampoule at relatively low temperature of $650^0C$ in order to reduce self intercalation of $VSe_2$. Single crystal flakes were obtained by vapor transport in over 15 days after keeping a pelletized $VSe_2$ at $700^0C$ inside a quartz tube. A large quantity of single crystals of different sizes (some of them as big as $10 \times 8$ $mm^2$), were obtained between 600 to $650^0C$ at the lower temperature end of the tube. The Rietveld refinement of the X-ray diffraction data of the powder sample done using Philips X'pert-PRO Diffractometer, shows the compound to form in the single trigonal phase (Space group: P$\bar{3}$m1) with the lattice parameters a = 3.356Å and c = 6.104 Å.

The electrical resistivity, thermoelectric power and Hall coefficient were measured from 10-300K using a APD Close cycle Helium refrigerator unit. Specific heat of the compound was measured from 2-300K by relaxation method using a Quantum Design PPMS (Physical Properties Measurement System).

## 3. Results and discussion

### 3.1 Thermoelectric Power

The thermoelectric power of the $VSe_2$ single crystal as well as the polycrystalline sample from 10-300K, is shown in the figure 1. The TEP of the compound shows a broad peak in temperature regime 70-110K below the CDW transition at 110K. Previous studies have attributed this region to be of incommensurate CDW 110K to commensurate CDW state at 70K [1, 2, 15, 16, 17]. This anomaly observed in TEP is sharper compared to that observed in the electrical resistivity and the magnetic susceptibility. Like the Hall coefficient, the CDW of the compound is more sensitive to the change in density of the states (DOS) at the Fermi surface, and show a sharp feature at the transition, where the carrier density changes.

The positive thermoelectric power of $VSe_2$ shows the dominance of the holes as the charge carriers. The band structure and Angle Resolve Photo Emission (ARPES) studies show the presence of the hole pockets at the center of Brillouin zone and the cylinder like electronic Fermi Surface at the zone corners [18, 19, 20]. However a positive thermoelectric power can also occur if the electrons of higher energy in the d band (which is narrow) of vanadium are scattered more (or have low smaller relaxation time), compared to the low energy electrons.

As seen from the figure, the TEP of the single crystal sample shows higher peak at the transition compared to the polycrystalline sample, and also the value of TEP at room temperature for single crystal was found to be larger by ~1µV/K. This difference in TEP values for single crystal and polycrystalline sample is rather very small and may vary with the sample quality. The TEP of both the single crystal and polycrystalline samples show linear temperature dependence (slope $dS/dT \sim 0.042 \mu V/K^2$) above the CDW transition. The TEP of $VSe_2$ measured down to 70K by Bruggen et al for their crystal flake and polycrystalline pellet also shows similar values at room temperature but shows a wave like behavior above the CDW transition [21]. This behavior cannot be accepted as there is no other transition reported in this temperature regime, so their result arises questions about the quality their sample. Compared to their data, our result measured on the good quality crystals shows sharper variation at CDW transition and linear temperature dependence is observed from 110K to 300K. We have already reported similar linear dependence of TEP down to 30K for the Fe intercalated $VSe_2$ single crystal, where the CDW transition gets suppressed for more than 5% Fe intercalation [12].

The linear temperature dependence for both thermoelectric power and electrical resistivity up to such a high temperature (300K) is quite unexpected for this metallic compound and indicate towards the presence of only electron-phonon scattering. The absence of phonon-phonon scattering contribution in TEP and resistivity of the compound is quite remarkable. The similar linear temperature dependences of resistivity and TEP up to high temperatures have been observed for the High temperature cuprates superconductor along the $CuO_2$ plane in their normal region [22]. This may indicate toward the similar the non-Fermi liquid behavior in the $VSe_2$ which is also a layered system.

*3.2 Specific Heat*

The temperature dependence of the specific heat (heat capacity) of the $VSe_2$ from 2-300K is shown in the figure 2, which shows a weak but noticeable anomaly at the CDW transition. It should be mentioned here that this anomaly was not yet observed in earlier reports. The inset of the figure shows the enlarged view of the specific heat anomaly. From the data we were able to get the electronic and phonon contribution parameters.

The total heat capacity $C_p$, in the absence of any magnetic contribution is given by $C_p = C_e + C_{lat} = \gamma T + \beta T^3$, where $C_e$ and $C_p$ are the electronic and lattice contribution to the specific heat and $\gamma$ and $\beta$ are respective coefficients. The values of γ, and β as determined from the intercept and slope of the $C_p/T$ versus $T^2$ graph in the temperature below 7K, were found to be 7 mJ/mol-$K^2$ and 0.604 mJ/mol-$K^4$ respectively. In this low temperature range (T < $\theta_D$/40) $C_p$ can be believed to be following the Debye theory of specific heat. The values of $\gamma$, and $\beta$ are well within the range as observed for other transition metal chalcogenides [13].

The Debye temperature $\theta_D$ was calculated from the β using the relation

$$\beta = \frac{12\, n\, R\, \pi^4}{5\, \theta_D^3}$$

Where n is the number of atoms per molecules, R is the gas constant (8.31 J/mole-K). This gives a value of $\theta_D$ to be 213K, which agrees well with the value 220±5K determined by Kamarchuk et al using the Debye frequency calculation from the point contact spectroscopy [23].

## *3.3 Electrical Resistivity*

Resistivity of four different flakes of 1T-$VSe_2$ from 10K to 300K measured along the *ab*-plane of the flakes is shown on the log-log scale in the figure 3. All the flakes were obtained from the same batch, but show marked difference in their Residual Resistance Ratio (RRR) as 28, 16, 15, and 8 for the flake numbered as 1, 2, 3 and 4 respectively. The Residual Resistance Ratio (RRR), which is defined as the ratio of resistivity at 300K and the resistivity at 2K, is a measure of quality of the compound (Larger RRR value indicate a better quality of compound). Our RRR value of 28 is equal to that obtained by Toriumi et.al in their crystal, prepared using Vanadium wires, and is the highest for all reported VSe$_2$ flakes till now [2]. The resistivity of all the flakes decreases on cooling and shows the CDW transition in the temperature range 106-144K. The flake with RRR value 28 shows CDW onset temperature T$_0$ (measured from the point where the resistivity deviates from its linear behavior) 106K, whereas the flake with RRR value 8 has highest T$_0$ of 144K, which is the highest of all reported results.

It indicates that impurities and strains inside the crystals facilitate the lattice distortion (CDW transition) like the effect of pressure as shown by Friend *et.al.* in their study [14], where it has been shown that in the presence of hydrostatic pressure, CDW transition shifts to high temperatures. The strains/impurities inside the crystals give rise to the increased overlap of the vanadium's narrow d band with the s-p bands of selenium, and in turn increasing the CDW onset temperature.

The crystal flake marked as 1 in the figure 3, shows a hysteresis around the transition. Similar feature has also been observed by Disalvo and Waszczack, in their susceptibility measurement of 1T-VSe$_2$ around 145-150K [7]. However no such hysteresis is seen in other flakes of lower RRR value and also for the polycrystalline pellets. Similarly the crystal prepared using vanadium wires by Toriumi et al, and with the same RRR value of 28 as of our crystal, also showed no hysteresis [2]. It has been shown by Schneemayer and coworkers that the samples prepared at 650$^0$C, in presence of excess Se, form in proper stoichiometric composition [15] and any variation from these conditions, leads to the metal rich phase, with extra vanadium going to the van der Waals gap between the layers. This Hysteresis found in the resistivity of sample 1, may be due to the presence of extra vanadium, at intercalated sites, and shows resemblance with the hysteresis in resistivity of polycrystalline sample of more than 10% Fe intercalated VSe$_2$ [12]. In our opinion the hysteresis may be related to the higher temperature of preparation leading to the higher point defects since it has been observed for the crystal prepared at 750-650$^0$C [CSY, present study], and 700-630$^0$C [Disalvo et al 7] growth condition, but not for the crystals grown at lower temperature 650-500$^0$C [Levy et al, 3].

The compounds with high RRR value were found to have higher value of resistivity. Similar effect has also been observed by Toriumi et.al and Levy et al in their separate studies [2, 3], which is quite contradictory to the Matthiessen rule for the electrical resistivity of metals.

The low temperature resistivity behavior of the 1T-VSe$_2$ also shows different type of temperature dependence below 50K. Toriumi et al found $T^4$ and $T^3$, dependence in the temperature regime 5-15K and 15-40K, respectively. They argued that the $T^4$ dependence can be manifested into the typical phonon scattering induced $T^5$ dependence in the higher quality sample [2]. Bayard et.al found the dominance of electron-electron scattering ($T^2$ dependence) [16]. However In all of our samples ~$T^{3/2}$ dependence was observed (not shown here), irrespective of their different RRR values. These different temperature dependences of resistivity as well as higher value of resistivity for the cleanest crystal (of lowest RRR value) does imply that the strains/dislocations and impurities inside the crystal play a significant role. We believe that these strains also have some temperature dependence, and hence Mattheissen rule is not obeyed in these circumstances.

*3.4 Hall Coefficient*

Hall coefficient was measured on the 8 μm thick, crystal flake of VSe$_2$, in the magnetic field H = 8 kOe and the transverse current I = 10mA, and the data were taken, for all combination of field and current directions at every set temperature value. The results are plotted in the figure 4, along with the electrical conductivity. The Hall coefficient is negative in sign and increases on sharply at CDW transition on cooling, and shows an almost 10 times reduction in the number of conduction carriers (from $1.04 \times 10^{16}$ per c.c. at 140K to $1.48 \times 10^{15}$ per c.c.at 15K). The large changes in the Hall coefficient near CDW transition is similar to as observed for second and third row transition metals NbSe$_2$ and TaS$_2$ at CDW transition. The formation of CDW state leads to the opening of a energy gap at the Fermi surface and reduces the number of conduction carriers. The value of Hall coefficient was obtained as $4.2\times10^{-3}$ cm$^3$/coulomb at 15K temperature by us, which is higher but comparable to other reported the values of R$_H$ at 4.2K, viz. $1\times10^{-3}$ cm$^3$/coulomb by Thomson et al, $2\times10^{-3}$ cm$^3$/coulomb by Bayard et al, and $3\times10^{-3}$ cm$^3$/coulomb by Bruggen et al [1, 16, 24].

Like the other transition metal dichalcogenides, the strong temperature dependence of R$_H$ in VSe$_2$ can also be understood in terms of mixed conduction, where electrons on different parts of the Fermi surface have wave functions of different character (s-like, p-like, d-like etc.). The electronic structure calculations at room temperature shows the overlapping of the narrow d-band of vanadium with the s-p band originating from valence electrons of Se, which generates a small amount of hole pockets at the Fermi surface [18, 19, 20]. At high temperatures the Hall coefficient is dominated by the high mobility and low density holes and compensates the Hall contribution from the electrons. Thermoelectric power results also show the presence of holes as a conduction carrier. The occurrence of the positive sign of the Seebeck coefficient and negative sign for R$_H$ below the room temperature indicates towards the presence of mixed conduction carriers in the system. However it is not easy to answer and need to be explored further.

## 4. Conclusions

The Thermopower and the specific heat of the 1T-VSe$_2$ single crystal and polycrystalline samples are reported along with the electrical resistivity and the Hall coefficient. The CDW anomaly is observed in all the properties around 110K. The TEP is found to be positive whereas the Hall coefficient is negative, are suggestive of the mixed conduction of carriers by the strongly correlated s-p-d hybridized bands in the compound. The electronic and phonon contributions of the specific heat are obtained as 7 mJ/mole-$K^2$ and 0.604 mJ/mol-$K^4$ which are in agreement with that for the other transition metal dichalcogenide compounds. The linear dependence of the Thermopower and the electrical resistivity above 100K is worth to be noted, and indicate towards the possibility of non Femi liquid behavior. The transport properties of 1T-$VSe_2$ single crystals are very sensitive to the preparation conditions. The defects/strains inside the crystals seem to have some temperature dependence and so the low temperature electrical resistivity cannot be understood by just Matthiessen rule. The charged density wave transition is also not sacrosanct at 110K as observed in most of earlier studies, but depends on the quality of the sample. In our case we case CDW onset temperature $T_o$ at 106K to 144K for the crystal of varies RRR values of 8 and 28 respectively. The presence of different scattering and impurities seems to enhancing the overlapping of the bands and the CDW onset temperature. The occurrence of hysteresis in the electrical resistivity for the crystal with RRR value 28 is also a subject of further study.

**Acknowledgement**

CSY acknowledges Council of Scientific and Industrial Research (CSIR), India, for the Senior Research Fellowship.

**Figure 1**

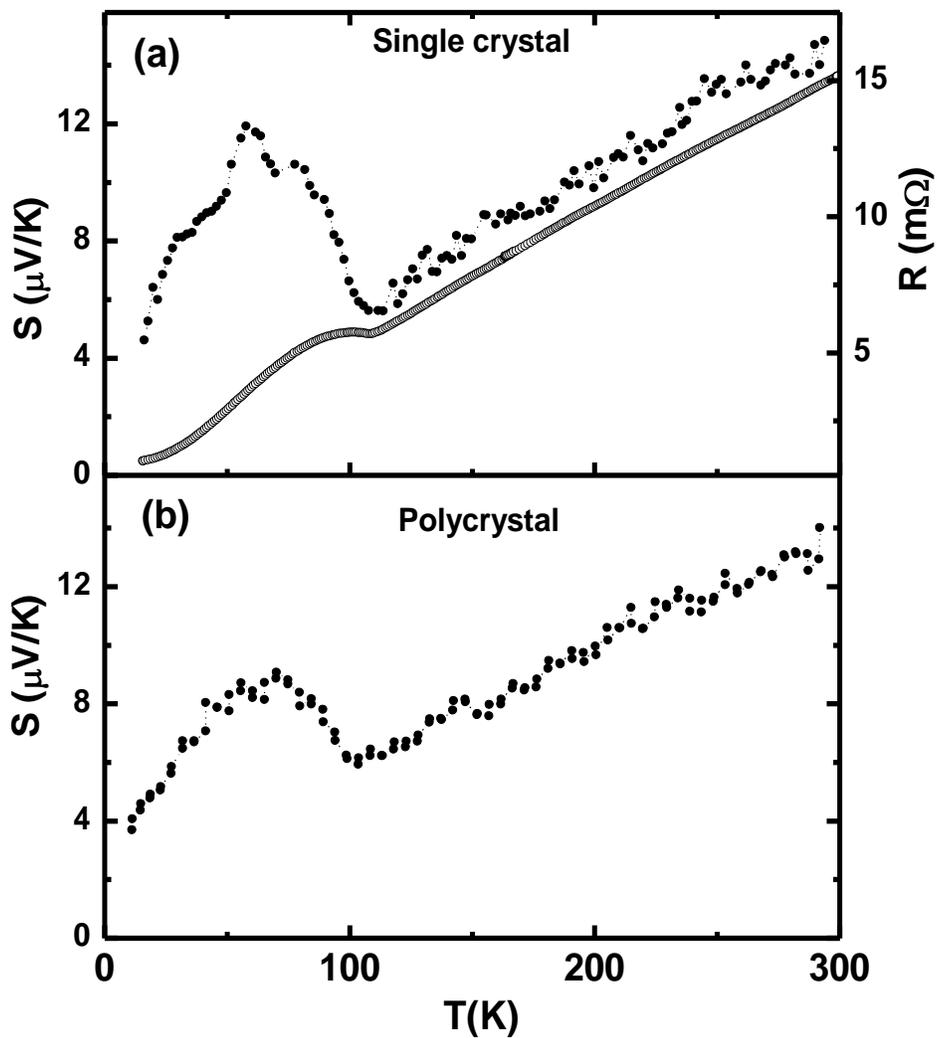

**Fig. 1.** The Thermoelectric power S of the single crystal (a) and polycrystalline (b) VSe$_2$ shows the linear temperature dependence above the CDW transition. The electrical resistivity for crystal is also plotted in figure 1(a) along with the TEP.

**Figure 2**

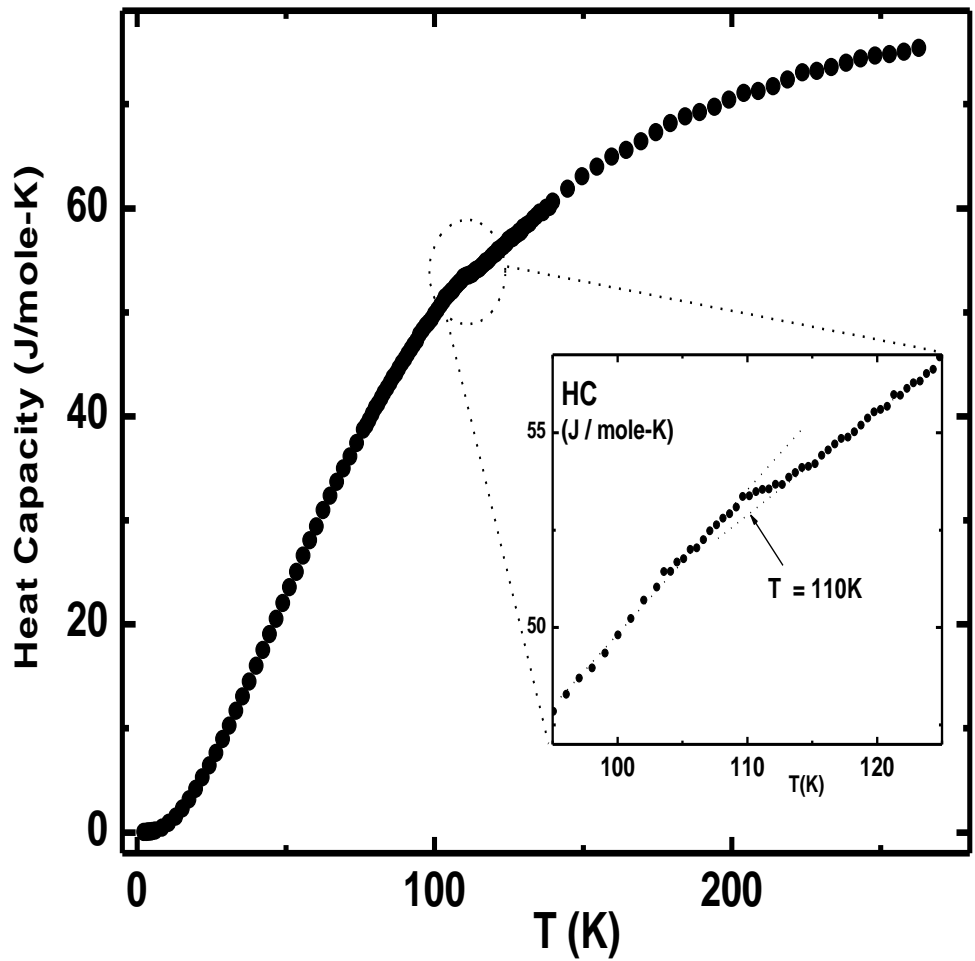

**Fig. 2.** The CDW anomaly is shown near 110K temperature in the $C_p/T$ versus T plot. Two straight lines are drawn to show the CDW clearly. The inset shows the same plot in temperature range 2-300K.

**Figure 3**

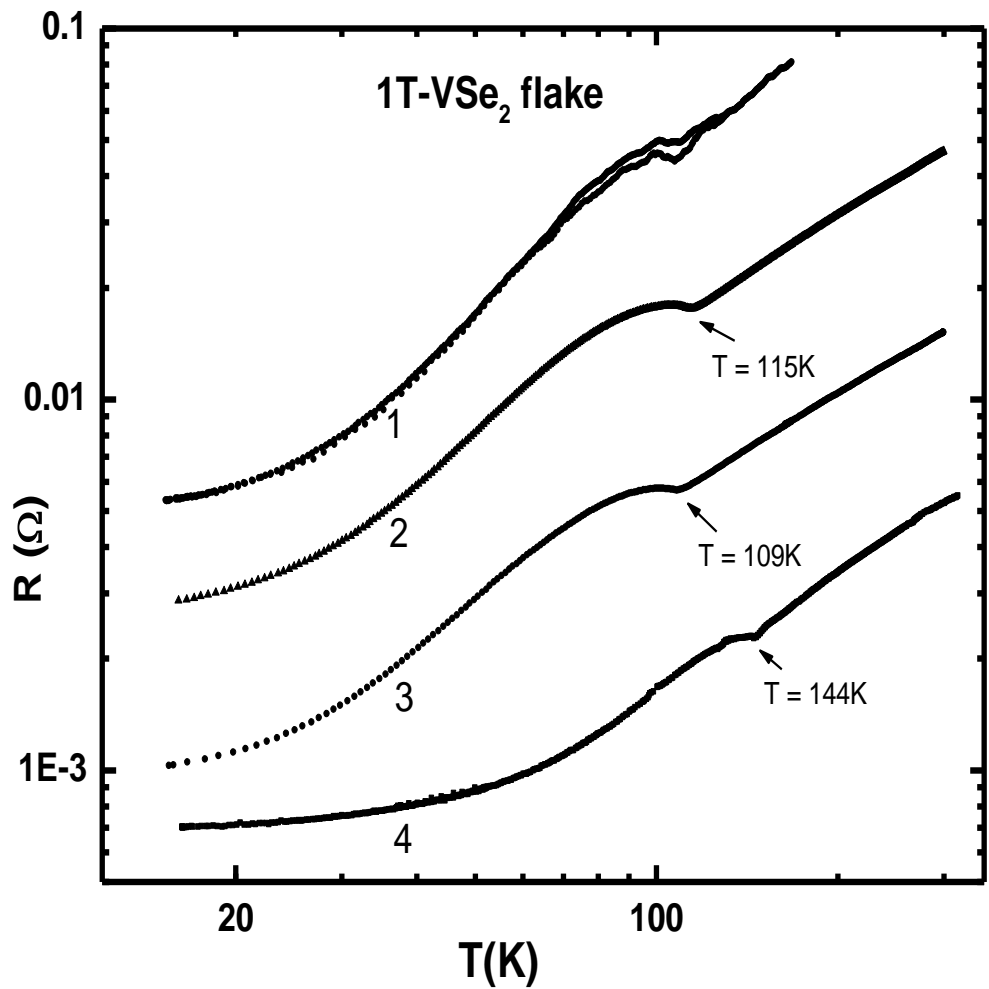

**Fig. 3.** Low temperature Resistance plots for the 1T-VSe$_2$ crystal flakes, plotted on the log-log scale to compare their low temperature behavior. Curve marked as 1, 2, 3 and 4 represents the flakes with RRR values 28, 16, 15 and 8 respectively.

**Figure 4**

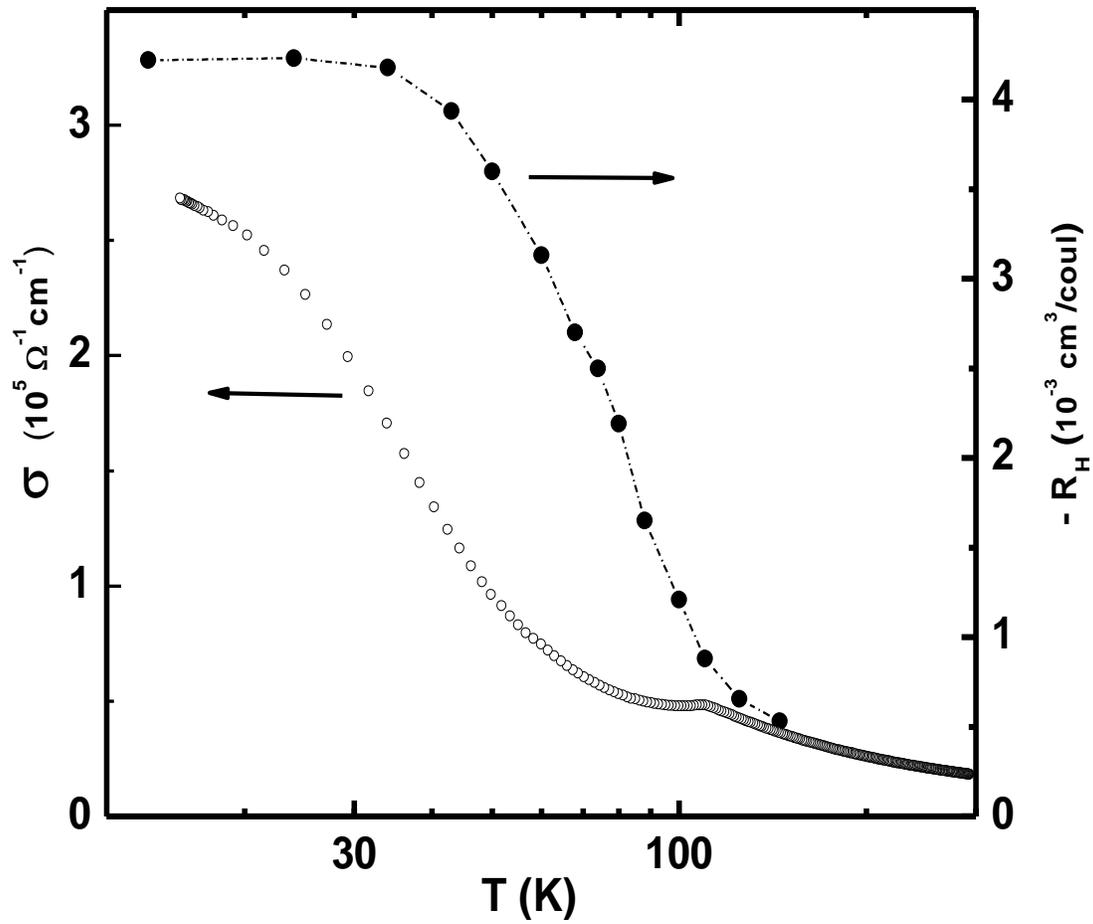

**Fig. 4.** The temperature dependence of the Hall coefficient of the 1T-VSe$_2$ is shown along with the electrical conductivity, measured on the crystal flake, shows a sharp variation at the CDW transition.

**Figure Caption**

**Fig. 1.** The Thermoelectric power S of the single crystal (a) and polycrystalline (b) VSe$_2$ shows the linear temperature dependence above the CDW transition. The electrical resistivity for crystal is also plotted in figure 1(a) along with the TEP.

**Fig. 2.** The CDW anomaly is shown near 110K temperature in the C$_p$/T versus T plot. Two straight lines are drawn to show the CDW clearly. The inset shows the same plot in temperature range 2-300K.

**Fig. 3.** Low temperature Resistance plots for the 1T-VSe$_2$ crystal flakes, plotted on the log-log scale to compare their low temperature behavior. Curve marked as 1, 2, 3 and 4 represents the flakes with RRR values 28, 16, 15 and 8 respectively.

**Fig. 4.** The temperature dependence of the Hall coefficient of the 1T-VSe$_2$ is shown along with the electrical conductivity, measured on the crystal flake, shows a sharp variation at the CDW transition